# In-plane and out of plane magnetic properties in $Ni_{46}Co_4Mn_{38}Sb_{12}$ Heusler alloys ribbons

Roshnee Sahoo[1], D. M. Raj Kumar[2], D. Arindha Babu[2], K. G. Suresh[1*] and M. Manivel Raja[2]

[1] *Department of Physics, Indian Institute of Technology Bombay, Mumbai- 400076, India*

[2] *Defence Metallurgical Research Laboratory, Hyderabad – 500058, India*

**Abstract:**

Magnetic, magnetocaloric and exchange bias properties have been systematically investigated in $Ni_{46}Co_4Mn_{38}Sb_{12}$ ribbon by applying magnetic field along (IP) and perpendicular (OP) to the ribbon plane. From the thermo-magnetization curves, the sharpness of the martensitic transition is observed to be nearly the same for both IP and OP ribbons. The thermomagnetic irreversibility region is found to be larger in the OP ribbon at 500 Oe, indicating that the magnetic anisotropy is larger in this case. The OP ribbon shows the Hopkinson maximum at 500 Oe, both for the FCC and ZFC modes. The magnetization curve for IP ribbon shows a faster approach to saturation, compared to the OP ribbon. Isothermal magnetic entropy change at 50 kOe has been found to be nearly same for both the ribbons. At 5 K the coercivity and exchange bias values are larger for the OP ribbon. Crystallographic texturing of the ribbons and its effect in the easy magnetization direction are found to be the reason behind the differences between the two ribbons.

*Corresponding author (email: suresh@phy.iitb.ac.in)

Associated with magneto-structural coupling near the martensitic transition, Heusler systems exhibit many interesting functional properties such as shape memory effect, magnetocaloric effect (MCE) and magnetoresistance (MR).[1-3] The magnetic properties in this system are mainly due to Mn magnetic moment as Ni moment is negligible. The Mn-Mn indirect exchange interaction (RKKY type) plays the key role behind the observed magnetic behavior in this system.[4] Off-stoichiometric composition of Ni-Mn-Sb Heusler systems have been drawing much interest recently due to the presence of magneto-structural transition and many interesting magnetic properties associated with this transition, which in certain compositions occurs near the room temperature. In the NiMnSb bulk system, MCE and MR properties has been studied extensively.[2,3,5-8] Recently melt spinning technique has gained much attention as it produces strongly textured polycrystalline ribbons. There are various reports on crystallographically textured melt spun ribbons where magnetic anisotropy plays the vital role in determining the observed properties. However, only very few reports deal with Heusler alloys as far as this aspect is concerned.[9] In the present work, we have studied the magnetic, magnetocaloric and exchange bias properties along two characteristic, in-plane (IP) and out of plane (OP), directions of the $Ni_{46}Co_4Mn_{38}Sb_{12}$ melt spun ribbon. This composition has been chosen as the bulk alloy of this composition has been reported to show martensitic transition near room temperature.[2]

Polycrystalline sample of $Ni_{46}Co_4Mn_{38}Sb_{12}$ was prepared by arc melting in high purity argon atmosphere. The constituent elements were at least of 99.99% purity. The ingot was remelted several times and the weight loss after the final melting was negligible. About 5 g of the as-melted alloy was taken in a quartz tube with a 1 mm diameter nozzle and was induction melted in flowing argon. The molten alloy was rapidly quenched in a high vacuum melt spinner, which yielded the as-spun ribbon samples of typically of 3-5 mm width and 20-30 μm thickness.

The magnetization measurements were carried out using a vibrating sample magnetometer attached to a Physical Property Measurement System (Quantum Design, PPMS-6500), where magnetic field up to 50 kOe is applied both in the plane of the film and perpendicular to the plane.

From the x-ray diffractogram (XRD) analysis it has been observed that the ribbon crystallizes in $B_2$ ordered phase at room temperature. Figure 1 shows the zero field cooled (ZFC) and the FCC (field cooled cooled) magnetization curves for the IP and OP ribbon samples for 500 Oe and 30 kOe. In the figure $A_S$, $A_F$, $M_S$ and $M_F$ denote the austenite and martensite transition temperatures. In the figures (a) and (b) the martensitic transition can be seen clearly for the IP and OP ribbons. Comparing the OP ribbon with the IP ribbon, at 500 Oe, it can be seen that the ZFC curve shows well defined maximum in both the cases, but the temperature corresponding to the maximum occurs at about 175 K for the IP ribbon while it occurs at about 225 K. It can also be seen that while the FCC curve in the OP ribbon shows a maximum (at the same temperature as that in the ZFC curve), there is no clear maximum in the case of the IP ribbon. The irreversibility between ZFC and FCC curves occurs over a larger temperature regime for OP ribbon in the martensite phase, which indicates larger anisotropy in this case. Furthermore, at 500 Oe, the OP ribbon shows a maximum (both in FCC and ZFC) just below the Curie temperature of the austenite phase ($T_C^A$). This kind of maximum is referred to as Hopkinson maximum and is usually seen in materials with fine grain structure.[9,10] Hernado et al.[9] have observed a similar effect in IP ribbons of NiMnSn Heuselr alloys[9]. These authors have also reported that the grain growth in these ribbons is normal to the ribbon plane and that the average grain size is much less than that of the bulk. However, it is of importance to note that in the present case, we do not see the Hopkinson effect in the IP ribbon, but only in the OP ribbon.

However, as the field is increased to 30 kOe, the behavior of the IP and the OP ribbons becomes quite identical with no anomalies (insets of Fig. 1). It can also be seen that for 500 Oe, the magnetization value at $M_S$ and $A_F$ is higher for the IP ribbon as compared to that of the OP ribbon. However, for 30 kOe, this trend reverses and the OP ribbon shows a slightly larger magnetization. These results signify the presence of larger anisotropy in the OP ribbon for low fields.

Figures 2(a) to (d) show the magnetic isotherms at different temperatures in the martensitic transition region for the IP and the OP ribbons. $H_1$ and $H_2$ as given in figure (c) denote the critical fields in the first loop where the slope changes for the IP and OP ribbons respectively. It can be observed that for 290 K, the magnetization increases continuously with increase in field. Here, the IP and OP critical fields merge with each other. This is expected as this temperature is close to the paramagnetic region of the martensite phase. In contrast, for 295 K, 300 K and 305 K, it can be seen that, with increase in field, there is a sudden slope change. After the critical fields $H_1$ and $H_2$, magnetization value increases continuously. At all these three temperatures, $H_2$ is found to be larger than $H_1$. Presence of larger OP anisotropy explains the observed anomaly. This fact is quite evident in the isotherms taken at 300 and 305 K. The observed hysteresis between increasing and decreasing fields at 295 and 300K is due to the metastability near the martensitic transition.

The isothermal magnetic entropy change $\Delta S_M (T, H)$ can be calculated by integrating the Maxwell relation.[11] Figure 3 shows the temperature variation of $\Delta S_M$ of IP and OP NiCoMnSb ribbons for different fields, evaluated from magnetic isotherms in the vicinity of the martensitic transition. It can be observed that both IP and OP curves give similar behavior with identical values. In both the cases, the entropy change is found to be positive, indicating the inverse magnetocaloric effect as in the bulk. The temperature corresponding to the maximum MCE values (~13 J/kg.K) is found to be 297 K. It may be noted that the entropy change for the bulk alloy of the same composition is 32 J/kg K.[12] It has also been observed that the average hysteresis loss is almost equal for both the IP and the OP ribbons. The small difference in magnetic anisotropy at these fields is responsible for this similarity. Therefore, in melt spun ribbon, the MCE properties do not depend on its crystallographic texture, at least in high fields.

In order to investigate the crystallographic texture effect on ferromagnetic (FM) and antiferromagnetic (AFM) interactions, exchange bias properties have been studied for both the IP and the OP ribbons. As given in figure 4(a), M-H data for the field range of -20 kOe to 20 kOe have been taken at 5 K in the field cooling (FC) mode with a cooling and measuring field of 50 kOe. The inset of 4(a) shows the zero field cooled data for the same measuring field. The double shifted loop in ZFC curves and negative shift FC curves clearly indicate the presence of exchange bias in the system, which arises due to the exchange anisotropy.[13] The main panel and the inset of figure 4(b) shows the coercive field ($H_C$) and the exchange bias field ($H_{EB}$) variations for different temperatures. It can be seen that for both IP and OP ribbons, $H_C$ first increases, reaches a maximum and then decreases with increase in temperature. As shown in the inset of figure 4(b), $H_{EB}$ decreases with increase in temperature. It can be seen that for 5 K, $H_C$ and $H_{EB}$ values for IP ribbon are 413 Oe and 374 Oe. The corresponding values for the OP ribbon are observed to be 486 Oe and 406 Oe. The larger OP anisotropy leads to higher $H_C$ and $H_{EB}$ values, as observed in other systems.[14,15] It is known that the exchange bias behavior mainly arises due to strong FM/AFM exchange coupling. The larger anisotropy of the OP ribbon than the IP ribbon may help in strongly coupling the AFM/FM layer at the interface, which may result in larger $H_{EB}$ for the OP ribbon. With increase in temperature, the thermal fluctuation causes a decrease in the exchange coupling. Therefore, both IP and OP $H_{EB}$ values decrease with increases in temperature. The blocking temperature for both IP and OP ribbons is 80 K where $H_{EB}$ is almost zero.

The observed anisotropy and distinct magnetic properties for OP ribbon clearly reveal the presence of crystallographic texture. In the rapid quenching process in the melt spun ribbon, the grains are very fine and have preferred orientation along a particular axis. Increase in the magnetization arises from the domain wall motion and the domain rotation. The former needs relatively less energy, especially in multi-domain particles. If this contribution is significant, the Hopkinson peak will not be very sharp. On the other hand, if the domain wall rotation is more dominant, the Hopkinson peak will be very sharp. If the ribbon is textured, applying field parallel to the ribbon plane would cause more grains to align in that direction. If the field is applied in other directions, then misalignment of grains away from the easy axis occurs only at higher energies, i.e., only at temperatures just below $T_C$.

In conclusion, the magnetic, magnetocaloric and exchange bias properties have been investigated for both IP and OP ribbons. From thermomagnetic curves it has been found that at 500 Oe, significant amount of anisotropy is present in the OP ribbon, which gets diminished at a field of 30 kOe. The OP ribbon shows the Hopkinson maximum at 500 Oe, both for the FCC and ZFC modes. Faster approach to saturation has been found for IP as compared to OP ribbon in magnetic isotherms taken near the martensitic transition. The critical field for OP ribbon where magnetization slope changes is always found to be larger than IP ribbon. Near martensitic transition temperature the difference in critical fields in OP and IP ribbons is larger. It has been found that the magnetic entropy change values for both IP and OP ribbon are nearly equal for the same field change, as the applied fields are larger than the anisotropy field. Both IP and OP ribbons show exchange bias properties. At 5 K, $H_{EB}$ values for IP and OP ribbons are 413 and 486 Oe respectively. The differences in the magnetic properties of IP and OP ribbons give a hint towards the direction of texturing in this system.

Acknowledgements

KGS thanks CSIR, Govt. of India for the financial support granted for carrying out this work.

**Figure captions:**

FIG.1. Temperature dependence of ZFC and FCC magnetization curves of $Ni_{46}Co_4Mn_{38}Sb_{12}$ ribbon (a) IP at H=500 Oe, Inset: IP at H=30 kOe, (b) OP at H=500 Oe, Inset: OP at H=30 kOe. Here IP refers to the case where the field is applied parallel to the ribbon plane and OP refers to the field applied perpendicular to the ribbon plane.

FIG. 2. Magnetic isotherms for IP and OP ribbons of $Ni_{46}Co_4Mn_{38}Sb_{12}$ at different temperatures near the martensitic transition region. Insets show the 1$^{st}$ quadrant isotherms up to 50 kOe.

FIG. 3. Temperature variation of isothermal magnetic entropy change ($\Delta S_M$) in IP and OP ribbons for different fields.

FIG.4. (a) FC (at 50 kOe) hysteresis loops of IP and OP ribbons at 5 K; Inset: ZFC hysteresis loops at the same temperature, (b) temperature variation of coercive fields ($H_C$) of the IP and OP ribbons; inset: exchange bias field ($H_{EB}$) at different temperatures for the IP and the OP ribbons.

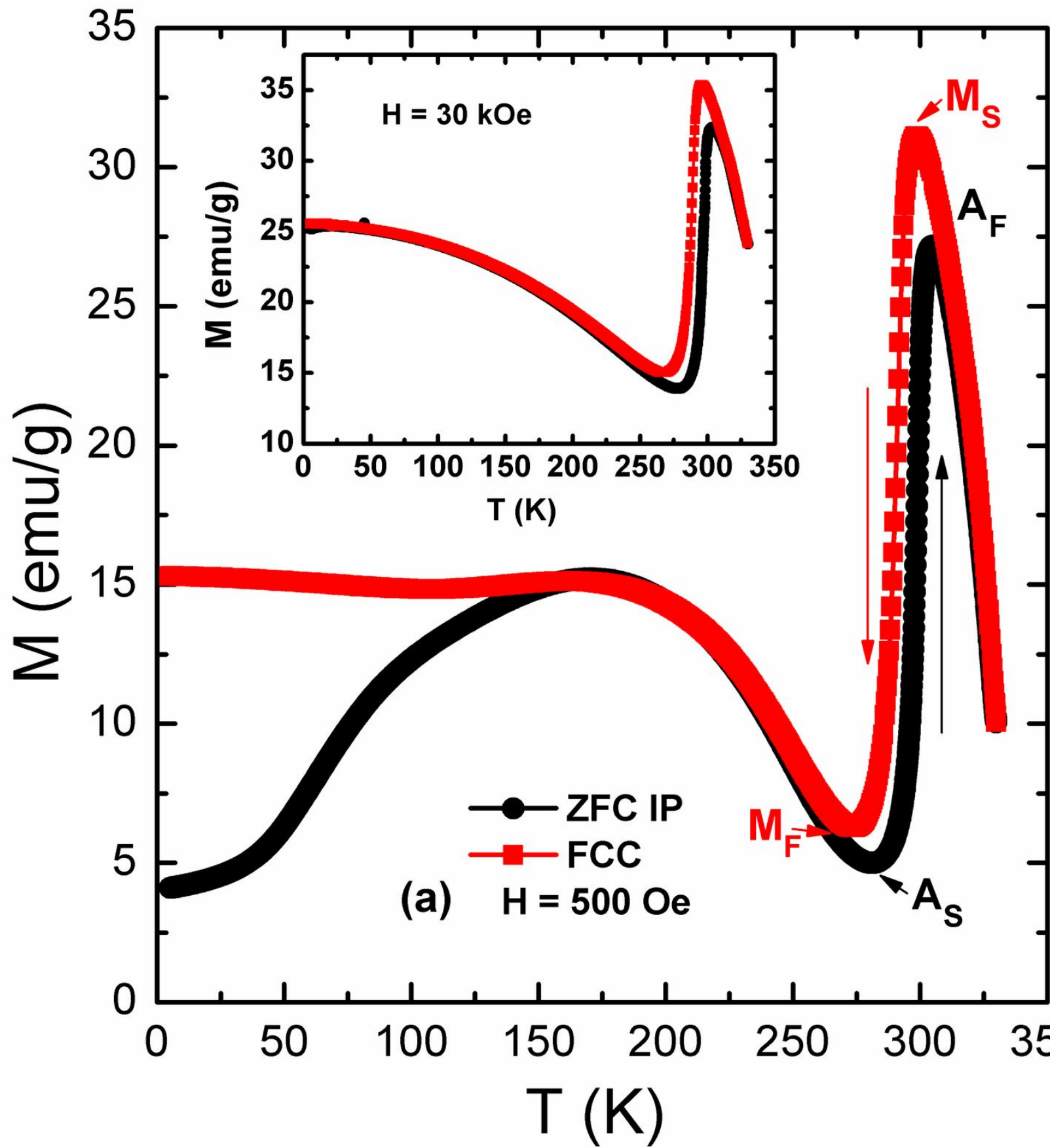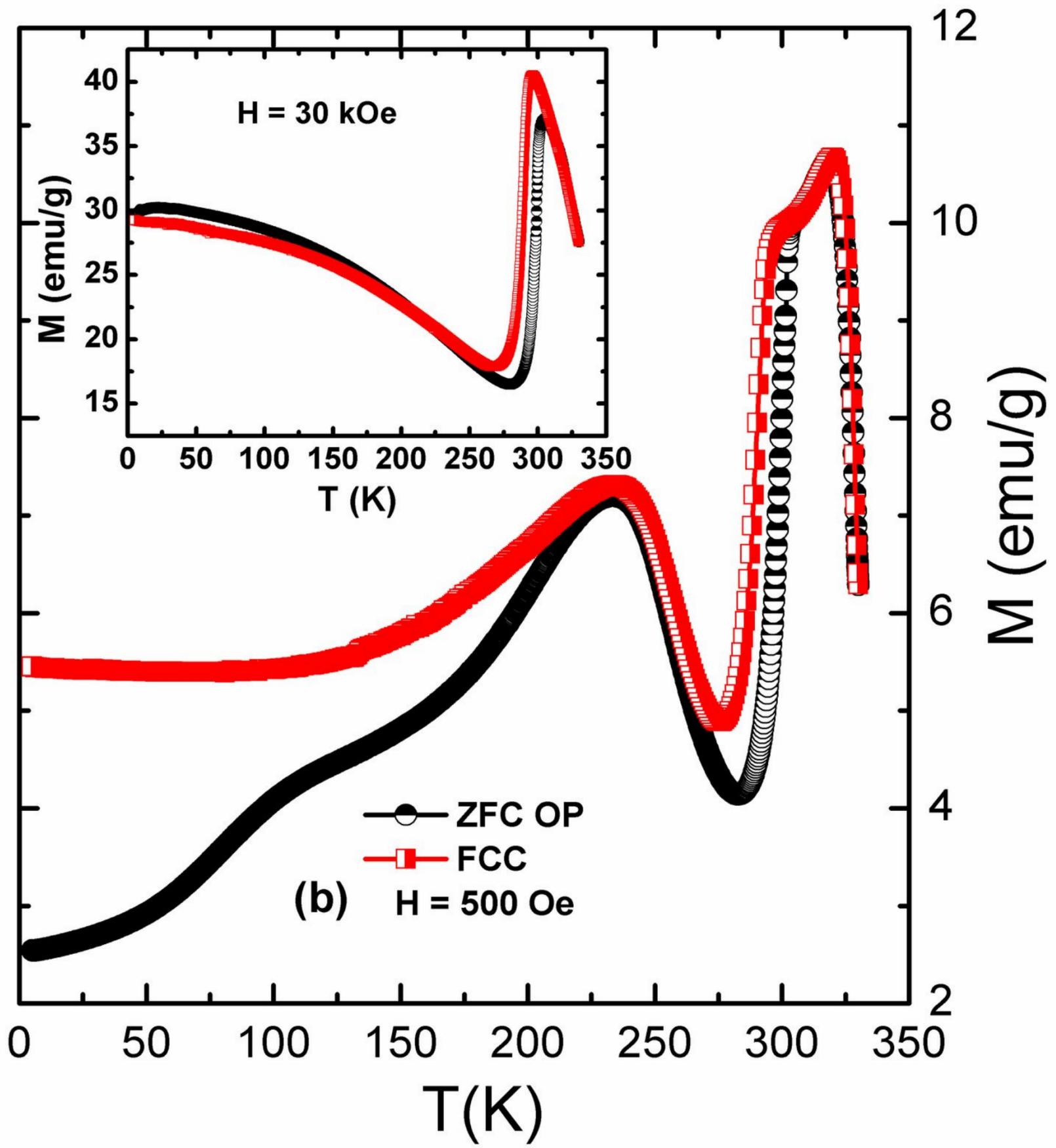

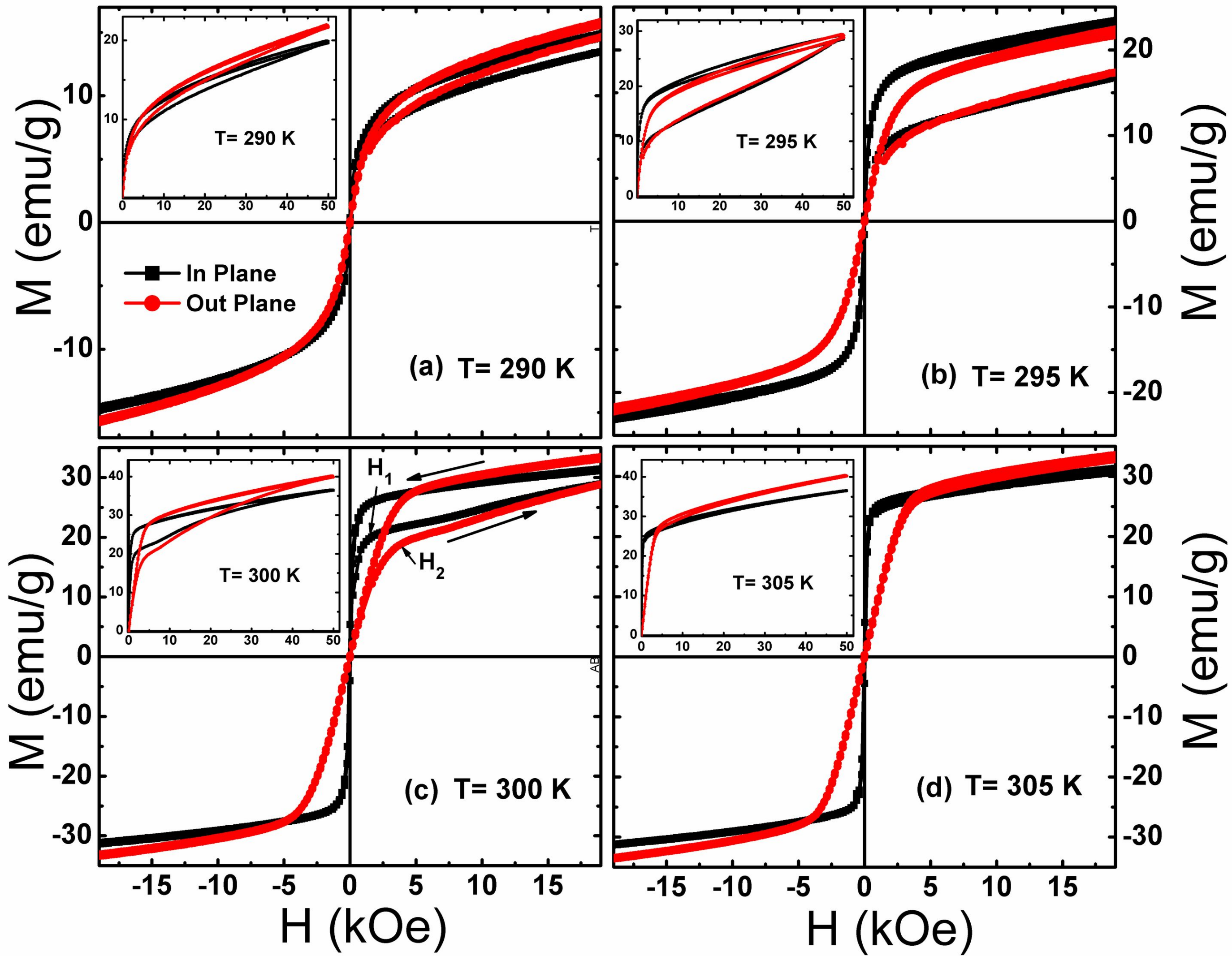

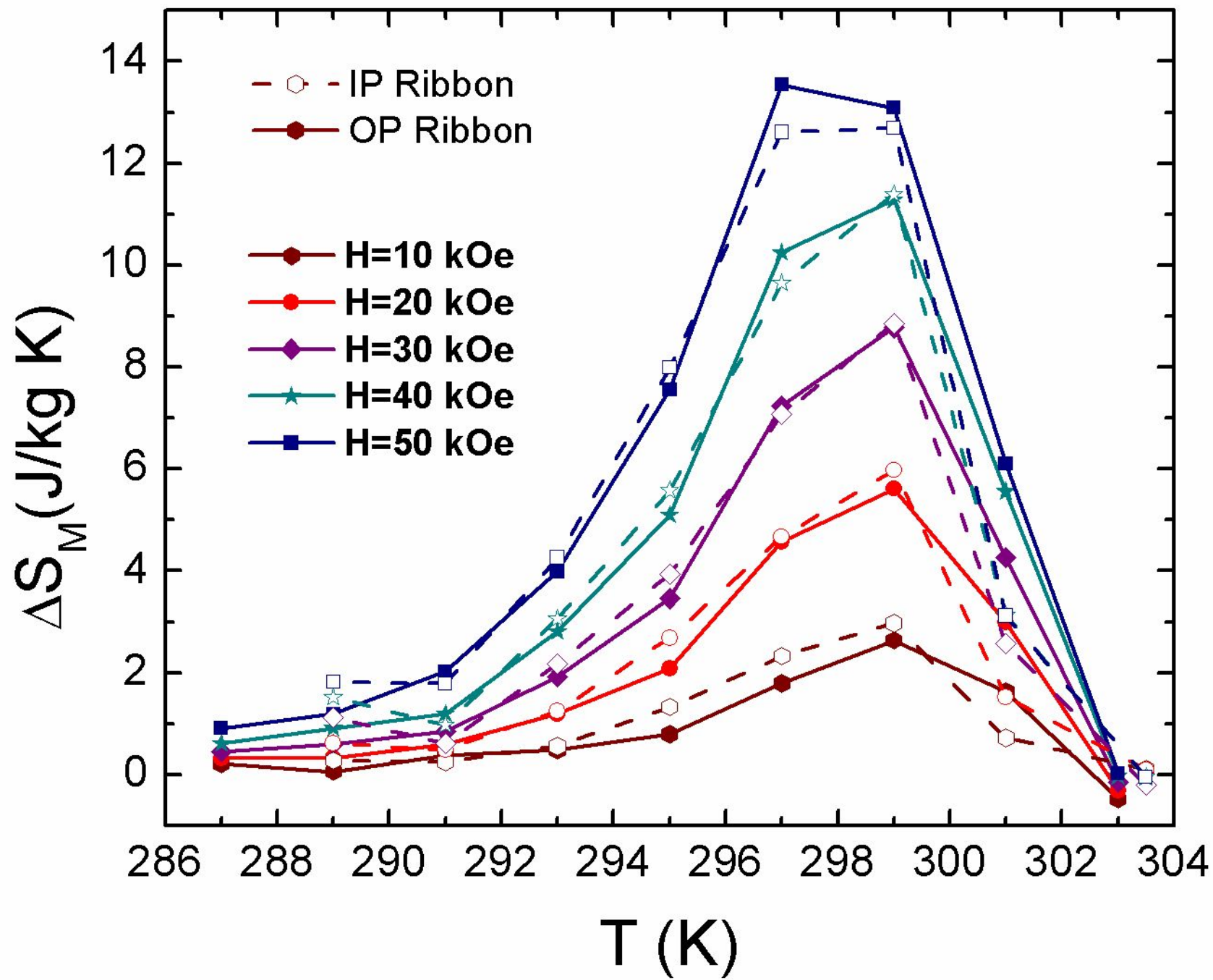

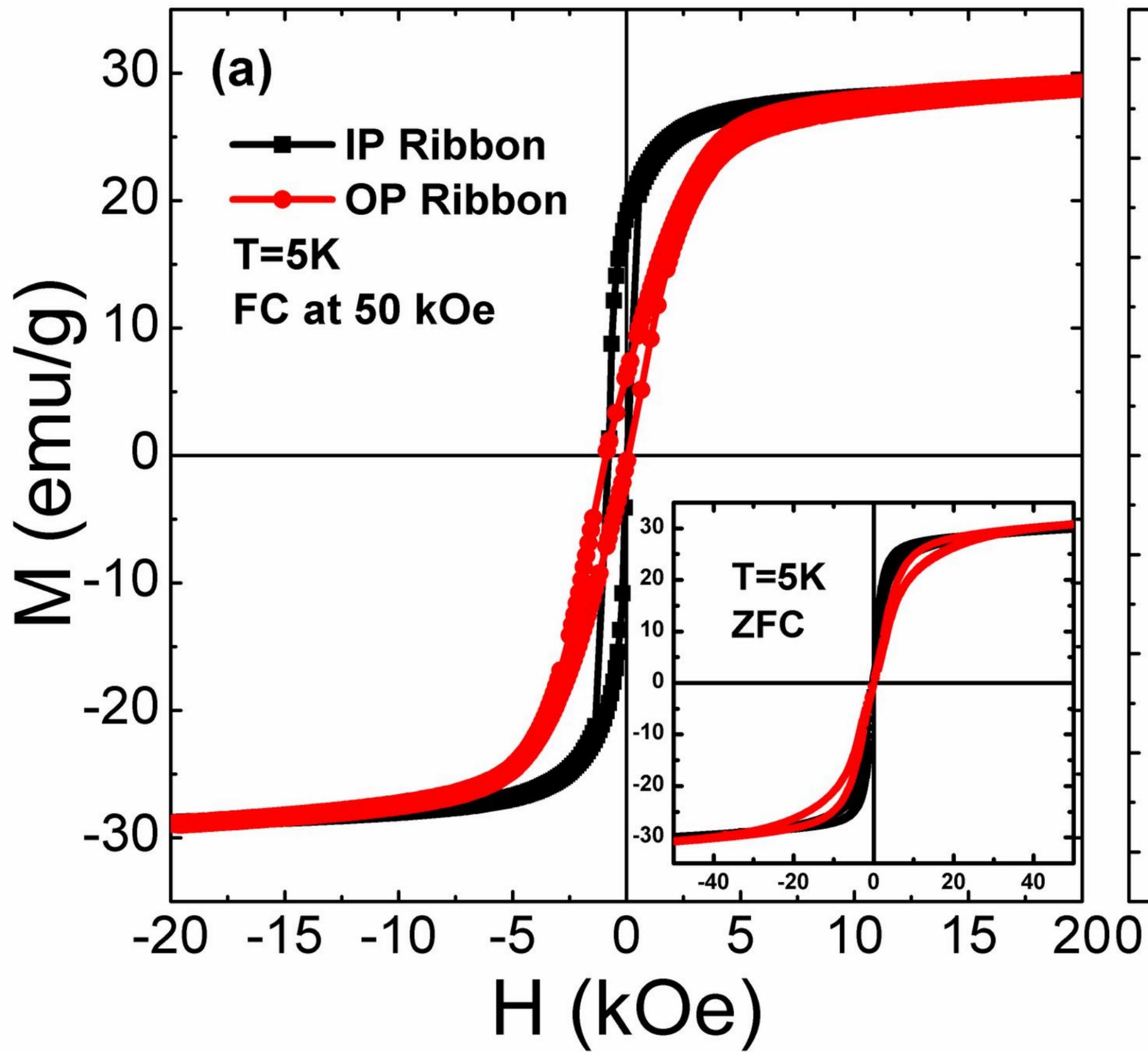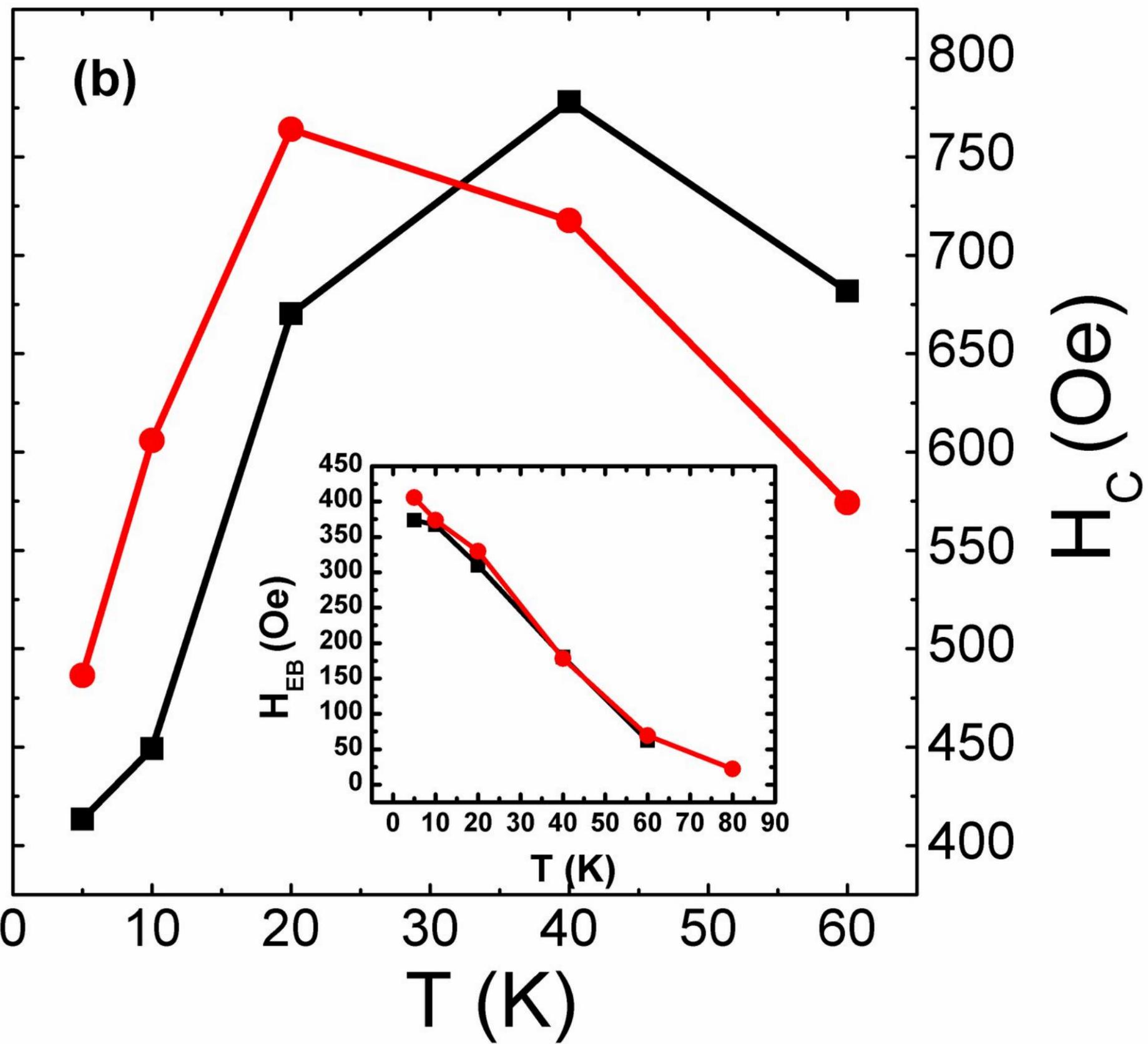